\documentclass[preprint,12pt]{elsarticle}%
\usepackage{epsfig}
\usepackage{amsmath}
\usepackage{graphicx,color}
\usepackage{amsmath}
\usepackage{amsfonts}
\usepackage{amssymb}%
\begin{document}
\begin{frontmatter}
\title{
 Geometrical Scaling and the Dependence \\of the Average Transverse Momentum\\ on the 
Multiplicity and Energy for the ALICE Experiment
}
\author{Larry McLerran\fnref{fn1}}
\address{Physics Dept, Bdg. 510A, Brookhaven National Laboratory, Upton, NY-11973, USA,\\
RIKEN BNL Research Center, Bldg. 510A, Brookhaven National Laboratory, Upton, NY 11973, USA,\\
Physics Department, China Central Normal University, Wuhan, 430079, China.}
\author{Michal Praszalowicz\fnref{fn2}}
\address{M. Smoluchowski Institute of Physics, Jagiellonian University, \\
Reymonta 4, 30-059 Krakow, Poland.}
\fntext[fn1]{e-mail: {\tt mclerran@me.com}}
\fntext[fn2]{e-mail: {\tt michal@if.uj.edu.pl}}
\begin{abstract}
We review  the recent ALICE data on charged particle multiplicity in p-p collisions, and show that it
exhibits Geometrical Scaling
(GS) with energy dependence given with characteristic exponent $\lambda=0.22$. 
Next, starting from the
GS hypothesis and using results of the Color Glass Condensate effective theory,
we calculate $\left\langle p_{\text{T}}\right\rangle$ as a function $N_{\rm ch}$
including dependence on the scattering energy $W$.
We show that $\left\langle p_{\text{T}}\right\rangle$  both in p-p and p-Pb collisions
scales in terms of scaling variable $(W/W_{0})^{\lambda/(2+\lambda)}%
\sqrt{N_{\mathrm{ch}}/S_{\bot}}$ where $S_{\bot}$ is multiplicity dependent
interaction area in the transverse plane.
Furthermore, we discuss how the behavior of the interaction radius $R$ at
large multiplicities affects the mean $p_{\mathrm{T}}$ dependence on $N_{\rm ch}$, 
and make a prediction that $\left\langle p_{\text{T}}\right\rangle$ at high multiplicity 
should reach an energy independent limit.
\end{abstract}
\end{frontmatter}

\setcounter{footnote}{0}

Recently, the  ALICE Collaboration has published charged particle spectra from p-p
collisions at three LHC energies 0.9, 2.76 and 7 TeV. This data, together with
p-Pb and Pb-Pb data has been subsequently used to construct other quantities,
in particular total multiplicity and mean $p_{\mathrm{T}}$ as a function of
charged particle multiplicity $N_{\mathrm{ch}}$
\cite{Abelev:2013ala,Abelev:2013bla}. This dependence has been next tested
against hypothesis of Geometrical Scaling (GS) proposed in
Ref.~\cite{McLerran:2013oju} with a conclusion that ``ALICE p-p and p-Pb data
at low and intermediate multiplicities are compatible with the proposed
scaling'' with substantial departure from scaling at larger multiplicities
\cite{Abelev:2013bla}. In this letter we show that GS works in fact much
better, provided one takes into account energy dependence of the interaction radius at fixed multiplicity.

The paper is organized as follows. First we introduce the basic concepts of
Geometrical Scaling for $p_{\mathrm{T}}$ spectra and show that the ALICE data
exhibit GS over a limited rage of $p_{\mathrm{T}}$ with energy dependence determined by the characteristic
exponent $\lambda= 0.22$.  This value  is slightly lower than the one found from the analysis
\cite{McLerran:2010ex} of single non-diffractive CMS data \cite{CMS}, and the
one found \cite{Praszalowicz:2012zh} in Deep Inelastic Scattering (DIS)
\cite{HERAdata}. Next, we derive the formula for mean transverse momentum and
discuss multiplicity dependence of the interaction radius. We use results of
the calculations \cite{Bzdak:2013zma} performed within the Color Glass Condensate
(CGC) effective theory \cite{MLV}. Finally we discuss the energy dependence of the
interaction radius and show that very good scaling of mean $p_{\mathrm{T}}$ is
seen in the ALICE data. We also argue that the character the energy dependence
changes for large multiplicities where the interaction radius schould reach energy
independent value. Such behavior has testable phenomenological consequences.
We finish with conclusions.

Multi-particle production at low and moderate transverse momenta probes the
nonperturbative regime of Quantum Chromodynamics (QCD). Yet at high energies
an overall picture drastically simplifies due to the existence of an
intermediate energy scale,  the saturation momentum $Q_{\mathrm{s}}(x)$.
Particle production proceeds via gluon scattering whose distribution is
determined by the ratio $p_{\mathrm{T}}/Q_{\mathrm{s}}(x)$ where Bjorken $x$
determines longitudinal gluon momentum. Therefore, on dimensional grounds,
the gluon multiplicity distribution is given in terms of the universal function
$\mathcal{F}(\tau)$ \cite{Kharzeev:2000ph,Kharzeev:2001gp,Kharzeev:2002ei}%
\begin{equation}
\frac{dN_{\rm g}}{dyd^{2}p_{\text{T}}}=S_{\bot}\mathcal{F}(\tau)\label{dNdydpT}%
\end{equation}
where%
\begin{equation}
\tau=\frac{p_{\text{T}}^{2}}{Q_{\text{s}}^{2}(x)}%
\end{equation}
is the scaling variable. Here
\begin{equation}
Q_{\text{s}}^{2}(x)=Q_{0}^{2} \left(  \frac{x_{0}}{x} \right)  ^{\lambda
}\label{Qsdef}%
\end{equation}
is the saturation momentum, $Q_{0}$ is an arbitrary scale parameter for
which we take 1 ~GeV$/c$, and for $x_0$ we take $10^{-3}$. 
Our analysis of GS presented in the present paper is not sensitive to the
actual value of $x_0$ and/or $Q_0$, but only to the value of $\lambda$.
Bjorken $x$'s of colliding partons for mid rapidity
production take the following form%
\begin{equation}
x=\frac{p_{\text{T}}}{W}.\label{Bjx}%
\end{equation}
Here $W=\sqrt{s}$ is the scattering energy 
and $S_{\bot}$ is a transverse area which will be specified later.
Equation~(\ref{dNdydpT}) exemplifies the property of the particle spectra known as
Geometrical Scaling (GS) where an observable that in principle depends on two
kinematical variables, such as $p_{\mathrm{T}}$ and $W$ (or Bjorken $x$),
depends in practice on a specific combination of them through the  scaling variable.
In Eq.~(\ref{dNdydpT}) we have suppressed strong coupling constant
$\alpha_{\mathrm{s}}$ whose dependence on $p_{\mathrm{T}}$ is expected to
introduce weak GS violation.


Equation (\ref{dNdydpT}) applies to the scattering of symmetric systems (pp, AA).
For pA scattering, which we will also discuss here, there are in principle two saturation scales:
the one of the proton $Q_{\rm s}^{({\rm p})}=Q_{\rm p}$, and that of the nucleus, $Q_{\rm s}^{(A)}=Q_{A}$, \cite{Kovchegov:1998bi,Dumitru:2001ux}. This issue has been 
discussed in Ref.~\cite{McLerran:2013oju} with a conclusion that for high enough multiplicities
and for central rapidities the two scales  should have the same energy dependence, meaning
that $Q_{\rm p}/Q_{A}={\rm const.}$ This condition is enough for GS of the form (\ref{dNdydpT})
to hold for pA collisions as well. A test of this assumption will be provided by the forthcoming 
pA data at a different LHC energy.


Geometrical scaling \cite{Stasto:2000er} has been introduced in the context of
DIS at HERA and later extended to particle production in hadronic collisions
\cite{McLerran:2010ex,Praszalowicz:2011tc,Praszalowicz:2011rm}. The saturation
scale appears due to the nonlinear effects in parton evolution with growing
energy. This evolution is in general described by the JIMWLK equation
\cite{jimwlk} which for large $N_{\mathrm{c}}$ reduces to the
Balitsky-Kovchegov equation \cite{BK}. These equations possess traveling wave
solutions which explicitly exhibit GS \cite{Munier:2003vc}.

A good description of large energy scattering, or equivalently of small
Bjorken $x$'s, is   the effective field
theory \cite{MLV} of the Color Glass Condensate (CGC) (for an introduction and review see
Ref.~\cite{Fukushima:2011ca}). In the theory of the CGC, hadrons after a collision stretch in the
longitudinal direction
strong gluonic fields that are coherent in the transverse plane over the
radius $1/Q_{\mathrm{s}}$. Multi-particle production proceeds by the decay of
these flux tubes, and it has been shown that the dominant contribution comes
from the production of gluons with $p_{\mathrm{T}} \le Q_{\mathrm{s}}$. This
mechanism is able to explain different features of high energy p-p collisions
including \emph{e.g.} negative binomial distribution \cite{Gelis:2009wh} or
ridge correlations in high multiplicity events \cite{Dumitru:2010iy}. In this
paper we shall use predictions of the CGC effective theory for the interaction
radii as functions of gluon multiplicity in p-p and p-Pb collisions discussed
in Ref.~\cite{Bzdak:2013zma}

\begin{figure}[h]
\centering
\includegraphics[height=7.0cm]{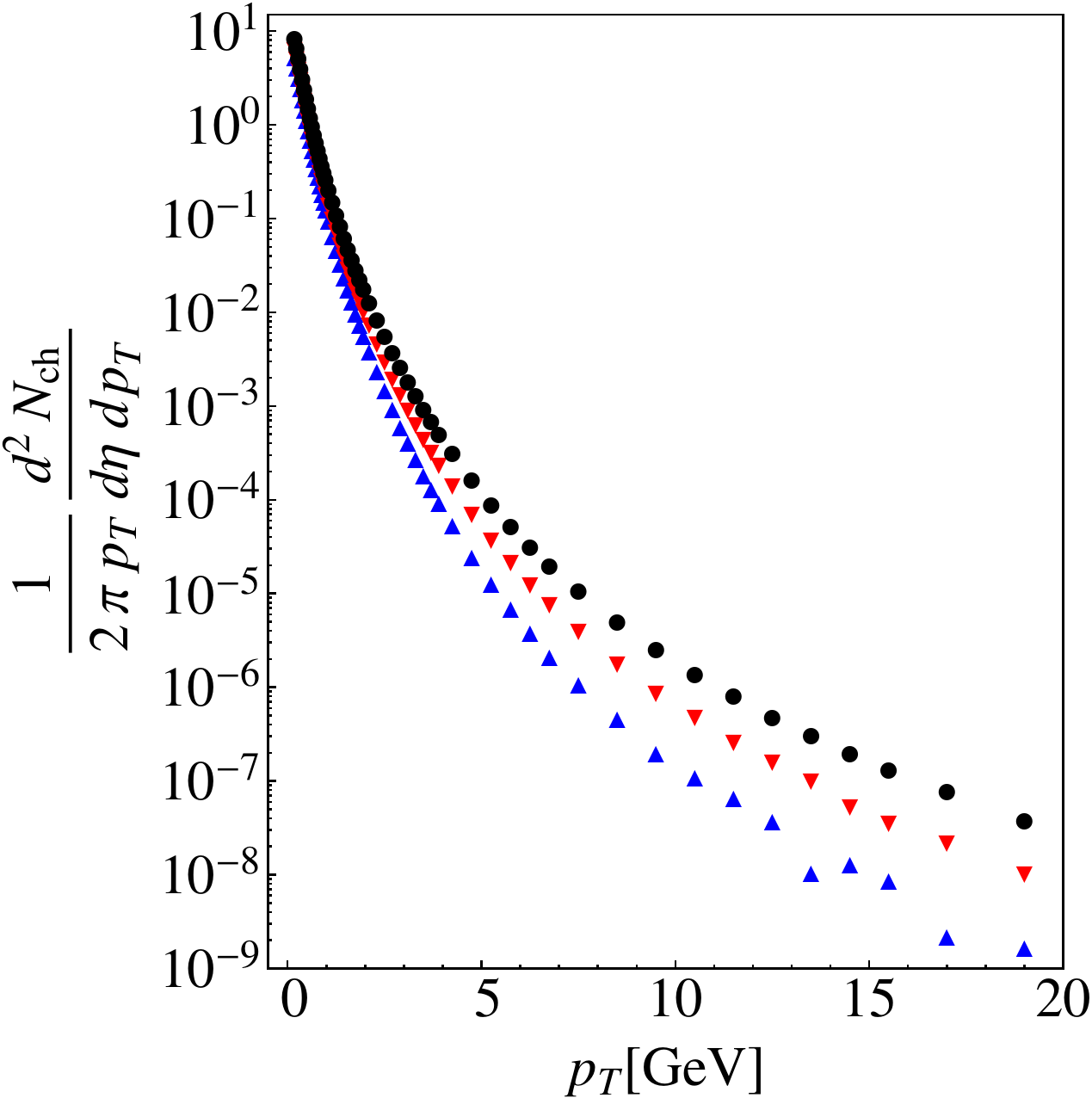}~~~~~~~
\includegraphics[height=7.0cm]{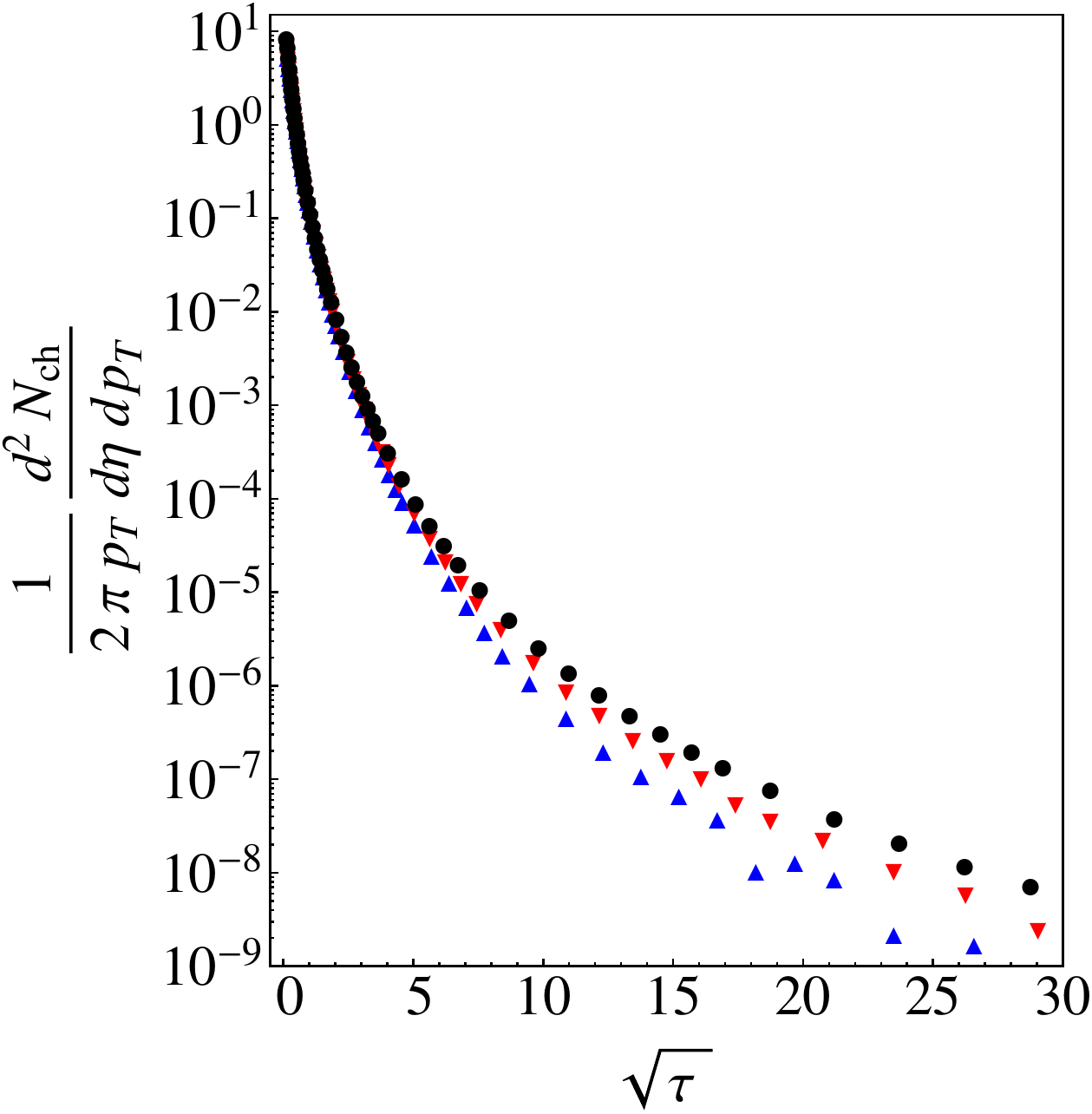}\caption{Multiplicity
distribution of charged particles in p-p collisions at 0.9 TeV (blue
up-triangles), 2.76 TeV (red down-triangles) and 7 TeV (black full circles)
plotted as functions of $p_{\mathrm{T}}$ (left) and as functions of the scaling
variable $\sqrt{\tau}$ for $\lambda=0.22$ (right).}%
\label{fig:spectra}%
\end{figure}

\begin{figure}[h]
\centering
\includegraphics[height=7.0cm]{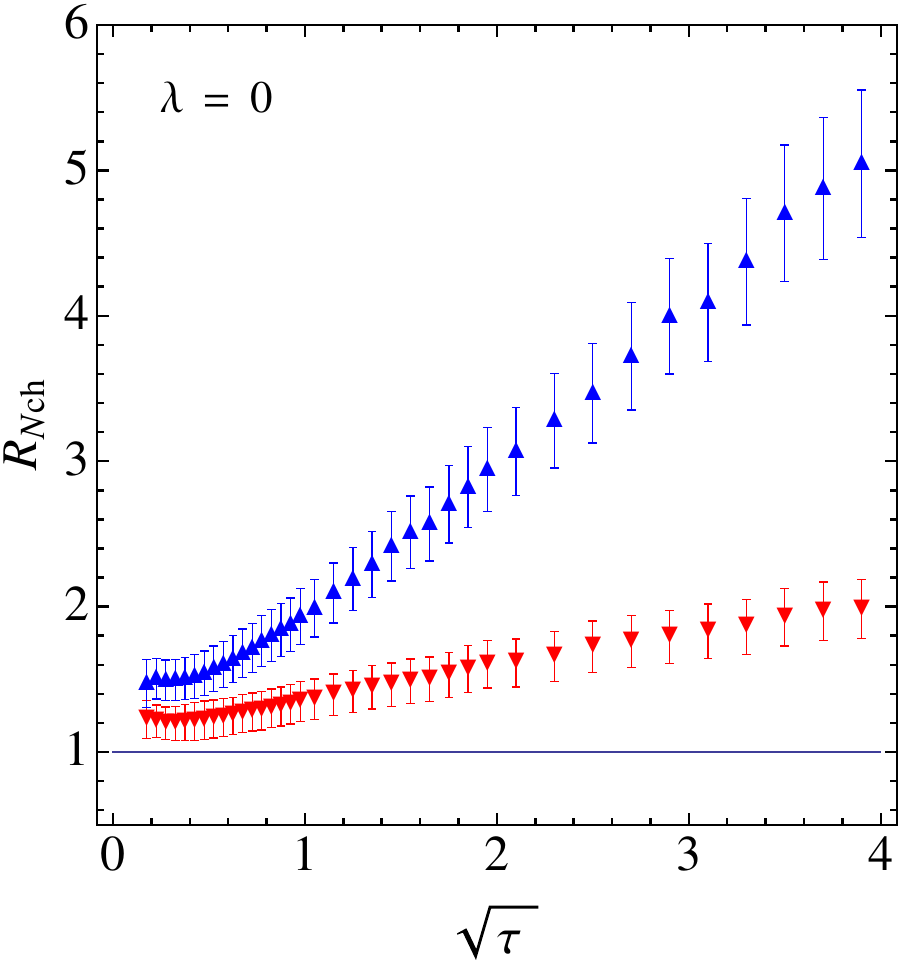}~~~~~~~
\includegraphics[height=7.0cm]{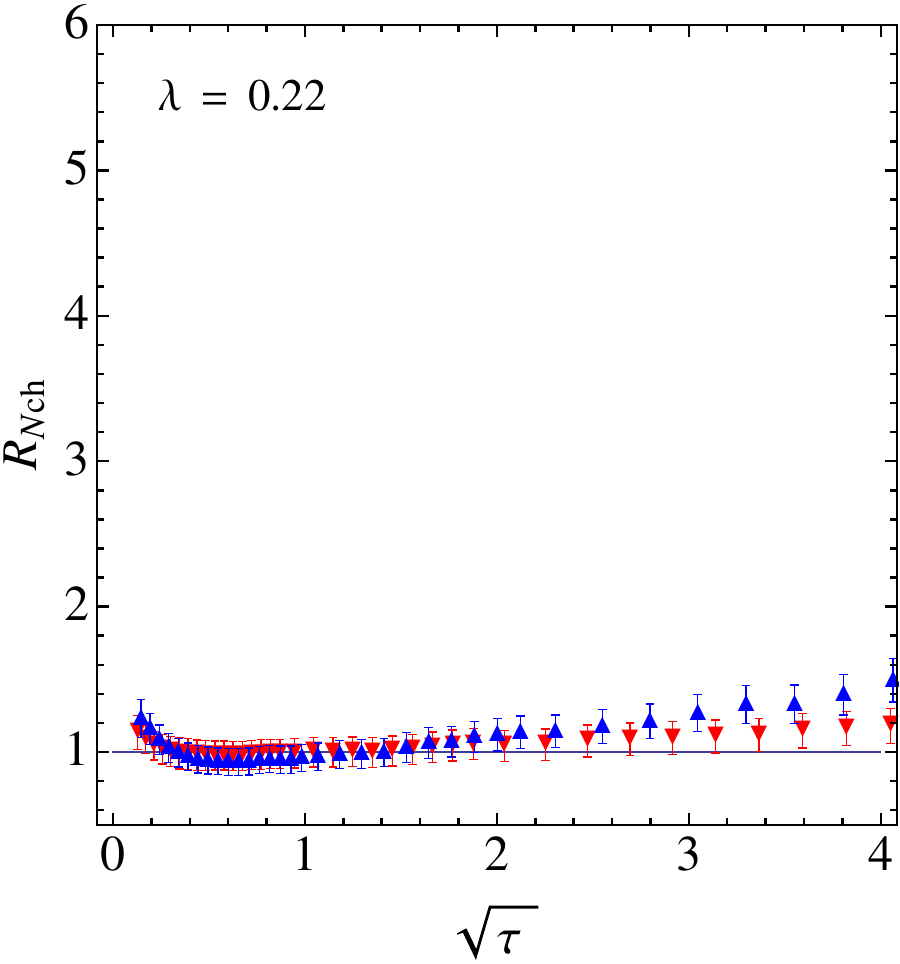}\caption{Ratios $\mathcal{R}%
_{12}$ (\emph{i.e.} 7 TeV to 2.76 TeV -- red down-triangles) and
$\mathcal{R}_{13}$ (\emph{i.e.} 7 TeV to 0.9 TeV -- blue up-triangles) plotted
as functions of $p_{\mathrm{T}}$ (left) and as functions of scaling variable
$\sqrt{\tau}$ for $\lambda=0.22$ (right).}%
\label{fig:ratios}%
\end{figure}

An immediate consequence of Eq.~(\ref{dNdydpT}) is that $p_{\mathrm{T}}$
spectra at different energies fall on one universal curve if plotted in terms
of the scaling variable $\tau$\footnote{In what follows we shall use $\sqrt{\tau}$
which for $\lambda=0$ reduces to $p_{\mathrm{T}}/Q_{0}$.}$^{)}$. The quality
of GS depends on the value of the exponent $\lambda$ entering the definition of the
saturation scale (\ref{Qsdef}). In order to determine $\lambda$ in a model
independent way we employ a method of ratios where we construct
\begin{equation}
\mathcal{R}_{ik}(\tau)=\frac{dN(W_{i},\tau)}{dyd^{2}p_{\text{T}}}/\frac
{dN(W_{k},\tau)}{dyd^{2}p_{\text{T}}}\label{Rik}%
\end{equation}
which, according to (\ref{dNdydpT}), should be equal to unity if GS is
present. In practice $R_{ik}\approx1$ in a window $\tau_{\mathrm{min}}%
<\tau<\tau_{\mathrm{max}}$. For particles of small $p_{\mathrm{T}}$
(\emph{i.e.} small $\tau$), comparable to $\Lambda_{\mathrm{QCD}}$ and/or pion
mass, we do not expect the arguments that lead to GS to be applicable, and for
large $p_{\mathrm{T}}$ we enter into a domain of large Bjorken $x$'s
(\ref{Bjx}) where GS is explicitly violated and perturbative QCD takes over.
In Eq.(\ref{Rik}) we have assumed that the number of charged particles is
proportional to the number of produced gluons and the proportionally factor
does not depend on energy (so called parton-hadron duality). We have checked
by explicit calculations of mean square deviations of $\mathcal{R}_{ik}$'s
from unity that the best value of $\lambda$ for the ALICE data that gives the
smallest $\chi^{2}$ over the largest interval in $\tau$ is equal to
0.22~\cite{Francuz}. This is illustrated in Fig.~\ref{fig:spectra} where in
the left panel we plot $dN/{dyd^{2}p_{\text{T}}}$ as functions of
$p_{\mathrm{T}} $ and as functions of $\sqrt{\tau}$ for $\lambda=0.22$ (right
panel). We see that spectra at different energies overlap within a window up
to $\sqrt{\tau}\sim4$. In order not to be biased by the logarithmic scale of
Fig.~\ref{fig:spectra}, we construct two ratios $\mathcal{R}_{12}$ and
$\mathcal{R}_{13}$ corresponding to the LHC energies $W_{1}=7$, $W_{2}=2.76$
and $W_{3}=0.9$~TeV, respectively. These ratios are plotted in
Fig.~\ref{fig:ratios} where again we plot them as functions of $p_{\mathrm{T}%
}$ (left panel) and as functions of $\sqrt{\tau}$ (right panel). We see
relatively good scaling where the weak rise $R_{ik}$'s with $\sqrt{\tau}$ can
be attributed to the residual dependence of $\lambda$ upon $p_{\mathrm{T}}%
^{2}$~\cite{Praszalowicz:2011tc}.

The behavior of ratios $\mathcal{R}_{ik}$ shown in Fig.~\ref{fig:ratios} is
almost identical as in the case of the CMS data analyzed in
Refs.~\cite{McLerran:2010ex}. However we have found in \cite{McLerran:2010ex}
that $\lambda_{\mathrm{CMS}}=0.27$ rather than $0.22$ and that the GS window
extends to slightly higher $\tau$. This may be due to the different event
selection (single non-diffractive at CMS vs. inelastic in ALICE) and different
pseudo rapidity coverage ($\left\vert \eta\right\vert <2.4$ at CMS vs.
$\left\vert \eta\right\vert <0.3$ at ALICE). In a recent study of GS in prompt
photon production \cite{Klein-Bosing:2014uaa} the optimal range of $\lambda$
turned out to be 0.22--0.28. These differences in $\lambda$ maybe in fact due
to some additional weak energy dependence of the multiplicity distribution
(\ref{dNdydpT}), like the one of $\alpha_{\mathrm{s}}$ or some energy
dependence of the unintegrated glue. 
Moreover one should bare in mind that there is no 
factorization theorem for multiparticle production in $k_{\rm T}$ dependent gluon 
density formalism. 
Studying ALICE data we have found for
example~\cite{Francuz}, that better GS quality is achieved for the
differential cross-section, rather than for the multiplicity, with
$\lambda\sim0.32$. Further discussion of these issues will be presented
elsewhere~\cite{Francuz}.

Having established the existence of GS in the ALICE data we can now proceed to
the analysis of total multiplicity and $\left\langle p_{\text{T}}\right\rangle
$. In order to calculate integrals over $p_{\text{T}}$ we need a Jacobian:%
\begin{align}
p_{\text{T}} &  =\bar{Q}_{\mathrm{s}}(W)\,\tau^{1/(2+\lambda)},\nonumber\\
dp_{\text{T}}^{2} &  =\frac{2}{2+\lambda}\bar{Q}_{\mathrm{s}}^{2}%
(W)\,\tau^{-\lambda/(2+\lambda)}d\tau
\label{pT2tau}%
\end{align}
where we have introduced an \emph{average} saturation scale%
\begin{equation}
\bar{Q}_{\mathrm{s}}(W)=Q_{0}\left(  \frac{W}{Q_{0}}\right)  ^{\lambda
/(2+\lambda)}
\label{Qbars}%
\end{equation}
which can be thought of as a solution of the equation%
\begin{equation}
Q_{\mathrm{s}}^{2}(\bar{Q}_{\mathrm{s}}/W)=\bar{Q}_{\mathrm{s}}^{2}.
\end{equation}
Note that due
to our choice of $x_0=10^{-3}$ in Eq.~(\ref{Qsdef}) $p_{\rm T}$ and $Q_0$ in
Eq.~(\ref{Qbars}) are in GeV and $W$ is in TeV. 

It follows that%
\begin{equation}
\frac{dN_{\rm g}}{dy}=A\,S_{\bot}\bar{Q}_{\mathrm{s}}^{2}(W)\label{dNoverdy}%
\end{equation}
where $A$ is an integral over the universal function $\mathcal{F}(\tau)$%
\begin{equation}
A=\frac{2\pi}{2+\lambda}%
{\displaystyle\int}
\tau^{-\frac{\lambda}{2+\lambda}}\mathcal{F}(\tau)d\tau
\end{equation}
and as such is energy independent. The constant $A$ can, however, depend on
particle species produced in the collision.

The universal  function $\mathcal{F}(\tau)$ is not known from first principles, however, in most
phenomenological applications where good description of the $p_{\rm T}$
spectra is given in terms of Tsallis parametrization, one can show that \cite{Praszalowicz:2013fsa}
\begin{equation}
\mathcal{F}(\tau) \sim \left(1+\frac{\tau^{1/(2+\lambda)}}{n \kappa}\right)^{-n}
\underset{\tau\rightarrow\infty}{\rightarrow} \left(\frac{n \kappa}{\tau^{1/(2+\lambda)}}\right)^n \, .
\label{Function}
\end{equation}
Here constant $\kappa$ follows from the fits to the transverse momentum spectra and
it is of the order of 0.1 \cite{McLerran:2013oju}, and power $n$ is of the order $5 - 9$.
In what follows we shall need this explicit form of $\mathcal{F}(\tau)$ only to estimate
possible energy dependence  of  its integrals, such as $A$ of Eq.~(\ref{dNoverdy}), coming from
the fact that experimentally $p_{\rm T}$ spectra are measured up to some  $p_{\rm T}^{\rm max}$
which translates into the maximal $\tau_{\rm max}$
\begin{equation}
\tau_{\rm max}=\left(\frac{p_{\rm T}^{\rm max}}{Q_0}\right)^{2+\lambda}
\left(\frac{Q_0}{W}\right)^\lambda ,
\label{taumax}
\end{equation}
which is energy-dependent. Here again $p_{\rm T}$ and $Q_0$ in
Eq.~(\ref{taumax}) are in GeV and $W$ is in TeV. It follows that for the ALICE data
$\tau_{\rm max} \approx 700$, 1300 and 1700 for scattering energies $W=0.9$, 2.76 and 7 TeV
respectively. Since $n$ is here numerically the highest power, the contributions of the unphysical tail, {\em i.e.}
from $\tau > \tau_{\rm max}$, are dumped approximately as $(1 / \tau_{\rm max})^{n/2-1}$
multiplied by a small coefficient $(n \kappa)^n$. Therefore in what
follows we shall neglect finite energy effects on the integrals of $\mathcal{F}(\tau)$. We have checked numerically
that the contribution to the mean transverse momentum coming from this effect is at the per mille level.

Equation (\ref{dNoverdy}) can be though of as a definition of the saturation scale,
which is essentially given as gluon number density per transverse area. One 
should however keep in mind that the transverse area itself does depend on
multiplicity as well, since it corresponds to the overlap area of two colliding
systems at fixed impact parameter $b$. The smaller $b$ the larger $S_{\bot}$.
This dependence will be of importance in the following, where we discuss
mean $p_{\rm T}$ dependence on multiplicity.

It follows from Eq.(\ref{dNoverdy}) that particle multiplicity in mid rapidity
grows like a power of the scattering energy, which is in remarkable
agreement with the LHC data~\cite{Aamodt:2009aa}. The power of this growth is
solely given by the energy rise of the average saturation scale $\bar
{Q}_{\mathrm{s}}^{2}$. Numerically for $\lambda=0.22$, we find that
$\lambda/(2+\lambda)\simeq0.099$, which again is in agreement with experimental
results~\cite{Aamodt:2009aa}. From this simple analysis we conclude that
$S_{\bot}$ is energy independent 
over the LHC energy range 
(or very weakly dependent for larger energy span).

When this formula is applied to minimum bias hadron-hadron collisions, we are basically fixing
the average hadron radius.  This average radius seems to be a slowly varying function of energy.
However, and this will be of primary importance in the following,
if we {\em fix} $dN_{\rm g}/{dy}$ and then {\em change} energy, then
$S_{\bot}$ has to change with energy as well in agreement with Eq.(\ref{dNoverdy}).  This is because
different radii are sampled at the different impact parameters.  
If we vary the density of particle per unit area, by varying the saturation momentum, and then require fixed multiplicity, 
we necessarily
will sample different impact parameters corresponding to different areas.

In heavy ion
collisions $S_{\bot}$ is equal to the geometrical transverse size of the
overlap of the colliding nuclei. As such it is related to the
centrality of the event and, in consequence, to the event multiplicity. It is
less clear what is geometrical interpretation of $S_{\bot}$ in p-p collisions. 
In a model with an impact parameter dependent saturation scale $Q_{\rm s}^b(b_\bot)$
\cite{Tribedy:2010ab} we have:
\begin{equation}
\bar{Q}_{\rm s}^2 S_{\bot}=\int d^2 r_{\bot}Q_{\rm s}^{b}(r_\bot)^2,
\end{equation}
where the integral extends over the overlap area  $S_{\bot}$ of colliding protons at a given
impact parameter $b$.
It is therefore obvious that also in the case of p-p scattering there should be a relation
between $S_{\bot}$ and multiplicity in a given event.  
Indeed, this dependence
has been calculated within the CGC framework~\cite{Bzdak:2013zma}, which
predicts that $S_{\bot}$ depends on $N_{\text{ch}}^{2/3}$ linearly, and then
saturates at some constant value. This behavior has a simple geometrical
interpretation: number of particles produced in hadronic collisions is
proportional to the active overlap volume. Once the maximal volume is reached,
further growth of multiplicity is due solely to fluctuations.

\begin{figure}[h]
\centering
\includegraphics[height=7.0cm]{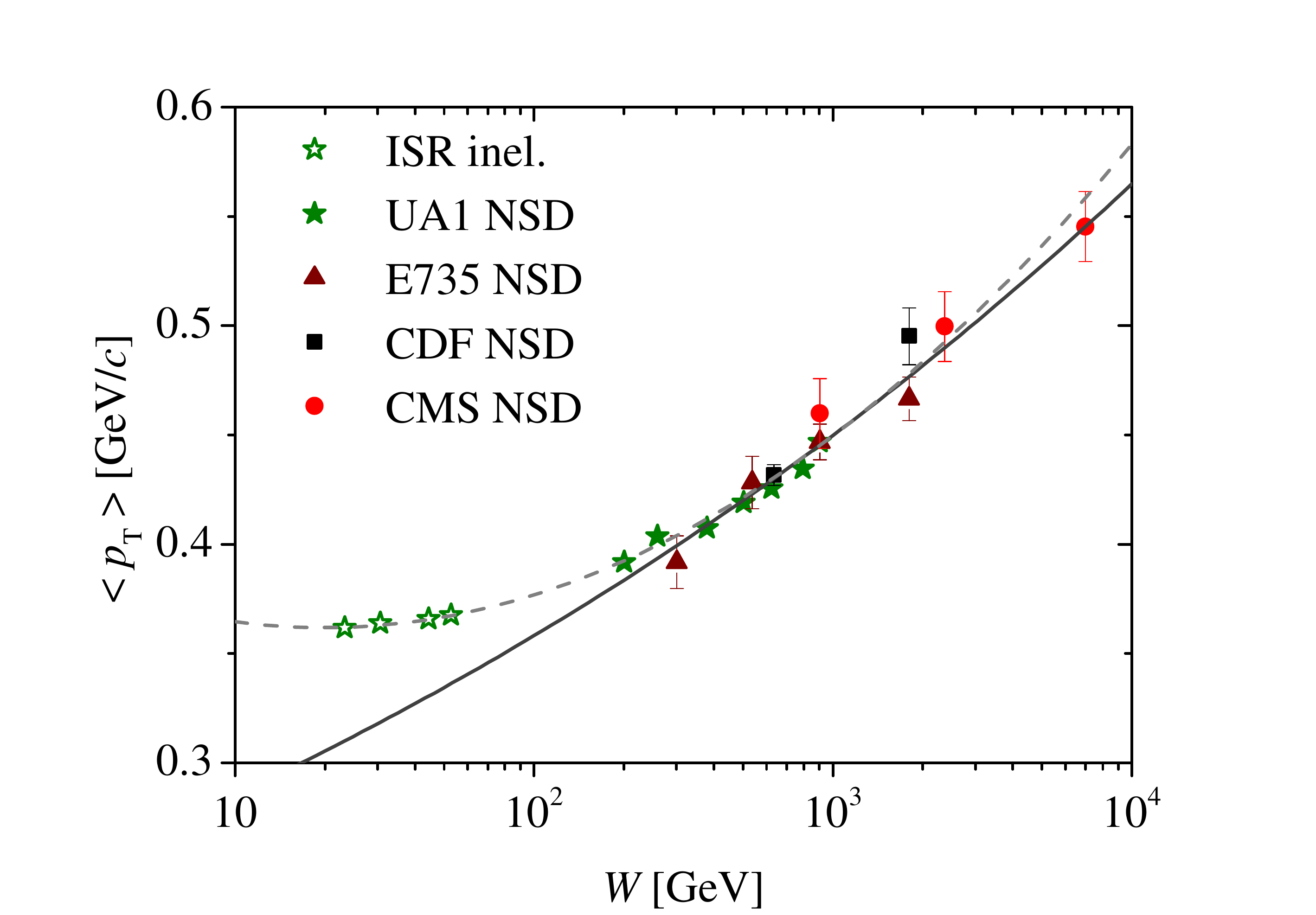}
\caption{Energy dependence of mean $p_{\rm T}$. Compilation of data from Ref.~\cite{Khachatryan:2010us}. Solid line
corresponds to the power law behavior of Eq.~(\ref{meanpT1}): $0.227\times W^{0.099}$, while the dashed line
corresponds to the CMS logarithmic fit aimed at describing also low energy data: 
$0.413-0.0171\,\ln s +0.00143\, \ln^2 s$, with $W^2=s$. }%
\label{fig:meanpTW}%
\end{figure}

Average $p_{\text{T}}$ can be easily calculated using Eqs.(\ref{dNdydpT}) and
(\ref{pT2tau}) giving%
\begin{equation}
\left\langle p_{\text{T}}\right\rangle =\frac{%
{\displaystyle\int}
p_{T}\frac{dN_{\rm g}}{dyd^{2}p_{\text{T}}}d^{2}p_{T}}{%
{\displaystyle\int}
\frac{dN_{\rm g}}{dyd^{2}p_{\text{T}}}d^{2}p_{T}}=\bar{Q}_{\mathrm{s}}(W)\frac
{B}{A}\label{meanpT1}%
\end{equation}
where%
\begin{equation}
B=\frac{2\pi}{2+\lambda}%
{\displaystyle\int}
d\tau\tau^{\frac{1-\lambda}{2+\lambda}}\mathcal{F}(\tau).
\end{equation}
Equation (\ref{meanpT1}) has two important consequences. First, it gives right
away the energy dependence of $\left\langle p_{\text{T}}\right\rangle $ which is illustrated
in Fig.~\ref{fig:meanpTW} where good agreement with the data taken from Ref.~\cite{Khachatryan:2010us}
can bee seen. Second,
at some fixed energy $W_{0}$ one can express $\bar{Q}_{\mathrm{s}}(W_{0})$ in
terms of the gluon multiplicity (\ref{dNoverdy}), which gives
\begin{equation}
\;\left.  \left\langle p_{\text{T}}\right\rangle \right\vert _{W_{0}}%
=\frac{B}{A}\, \sqrt{\frac{dN_{\rm g}/dy}{A\left.  S_{\bot}(dN_{\rm g}/dy)\right\vert _{W_{0}}}%
}.\label{pToneterm0}%
\end{equation}
Note that for fixed ${dN_{\rm g}/dy}$ interaction  size $S_{\bot}$ in Eq.(\ref{pToneterm0}) depends,
as explained above,  
on the reference energy $W_{0}$ and also on ${dN_{\rm g}/dy}$ itself,  which is
related to the number of charged particles $N_{\text{ch}}$ 
in the kinematical range of a given experiment:
\begin{equation}
N_{\text{ch}}=\frac{1}{\gamma\Delta y\,}%
{\displaystyle\int\limits_{\Delta y}}
\frac{dN_{\rm g}}{dy}dy\label{NgvsNch}.%
\end{equation}
Here the coefficient $\gamma$ relates gluon multiplicity to the multiplicity of
observed charged hadrons within the rapidity interval $\Delta y$. For ALICE
data used in this paper $\Delta y=0.6$. The interaction radius $R$ characterizing
the volume from which the particles are produced and which is related to
$S_{\bot}=\pi R^{2}$, depends in a natural way on the third root of
$dN_{\rm g}/dy$, \emph{i.e.} $R=R(\sqrt[3]{dN_{\rm g}/dy})=R(\sqrt[3]{\gamma
N_{\text{ch}}})$ \cite{Bzdak:2013zma} and on the collision energy $W_{0}$.

In the following we shall use a slightly modified formula for $\left\langle
p_{\text{T}}\right\rangle $, which takes into account nonperturbative effects
and contributions from the particle masses encoded in a constant $\alpha$:%
\begin{equation}
\left.  \left\langle p_{\text{T}}\right\rangle \right\vert _{W_{0}}%
=\alpha+\beta\,\frac{\sqrt{N_{\text{ch}}}}{\left.  R(\sqrt[3]{\gamma
N_{\text{ch}}})\right\vert _{W_{0}}}.\label{meanpTfit}%
\end{equation}
Here $\alpha$ and $\beta$ are constants that do not depend on energy. Formula
(\ref{meanpTfit}) has been proven to work very well in p-p at 7 TeV and also in p-Pb collisions
at 5.02 TeV at the LHC 
\cite{McLerran:2013oju}.

The interaction radius $R(\sqrt[3]{dN_{\rm g}/dy})$ has been calculated in
Ref.~\cite{Bzdak:2013zma}. Here we shall use the parametrization of
Ref.~\cite{McLerran:2013oju}:%
\begin{equation}
R_{\mathrm{pp}}(x)=\left\{
\begin{array}
[c]{ll}%
0.387+0.0335x+0.274\,x^{2}-0.0542\,x^{3} & \mbox{if $x < 3.4$,}\\
1.538 & \mbox{if $x \geq 3.4 $.}
\end{array}
\right.  [\text{fm}]\label{Rpp}%
\end{equation}
for p-p collisions at 7 TeV and%
\begin{equation}
R_{\mathrm{pPb}}(x)=\left\{
\begin{array}
[c]{ll}%
0.21+0.47\,x & \mbox{if $x < 3.5$,}\\
1.184-0.483\,x+0.305\,x^{2}-0.032\,x^{3} & \mbox{if $3.5 \leq x < 5 $,}\\
2.394 & \mbox{if $x \geq 5 $.}
\end{array}
\right.  [\text{fm}]\label{RpPb}%
\end{equation}
for p-Pb collisions at 5.02 TeV.

In our analysis of the 7 TeV ALICE data, it turns out we are only sensitive to radii
where  we are in the first interval for the dependence of $R$ on multiplicity.   This is the region
where impact parameter is varying.  In the very high multiplicity region, the radius saturates
and at lower energies it happens for lower multiplicities.

\begin{figure}[h]
\centering
\includegraphics[height=7.0cm]{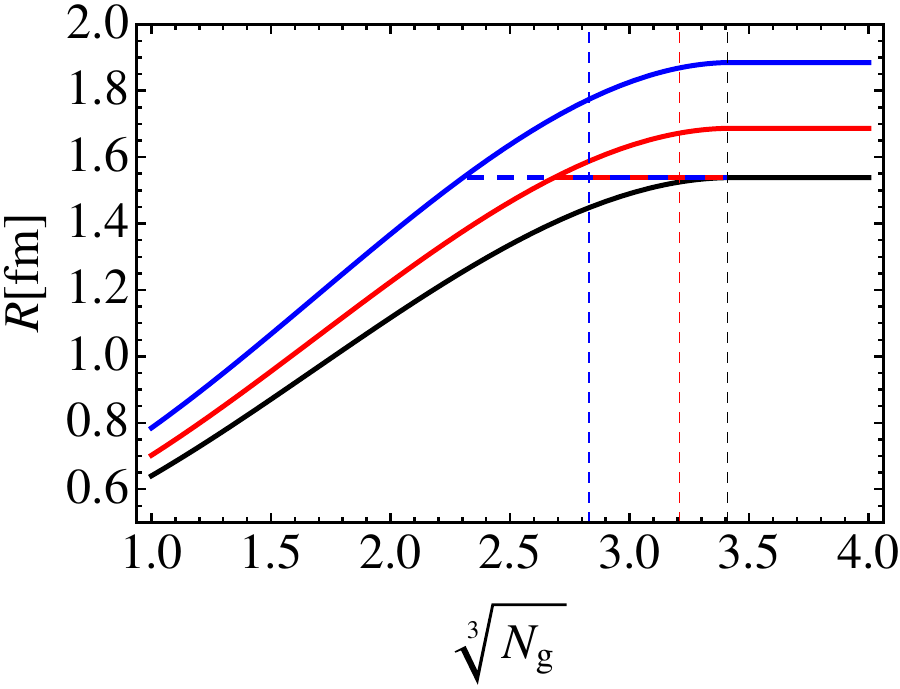}
\caption{$R(N_{\rm g}^{1/3})$ for three different energies.
The lowest solid (black) curve corresponds to the parametrization (\ref{Rpp}) at $W_0=7$~TeV, 
whereas two upper solid curves correspond to  (\ref{Rpp}) multiplied by the energy dependent factor
$(W_0/W)^{\lambda/(2+\lambda)}$ for $W=2.76$~TeV (red) and 0.9~TeV (blue). High multiplicity
saturation is schematically depicted by dashed lines. Vertical thin dashed lines correspond to
the highest multiplicities analyzed by the ALICE Collaboration at 0.9~TeV (most left blue), 2.76~TeV
(middle red) and 7~TeV (most right black).}%
\label{fig:radii}%
\end{figure}

The energy dependence of (\ref{meanpTfit}) follows from the general form given
by Eq.(\ref{meanpT1}). In order to find an explicit formula for 
mean $p_{\rm T}$ at any scattering energy $W$, {\em i.e} for
$\left. \left\langle p_{\text{T}}\right\rangle \right\vert _{W}$, 
one has to recompute
$R(\sqrt[3]{\gamma N_{\text{ch}}})$, but -- as a consequence of
Eq.(\ref{meanpT1}) -- one should obtain that%

\begin{equation}
\left.  \left\langle p_{\text{T}}\right\rangle \right\vert _{W}=
{\alpha}
+{\beta}\,\frac{\sqrt{N_{\text{ch}}}}{\left.  R(\sqrt[3]{{\gamma} N_{\text{ch}}%
})\right\vert _{W}}=
{\alpha}+{\beta}\,\left(  \frac{W}{W_{0}}\right)
^{\lambda/(2+\lambda)}\frac{\sqrt{N_{\text{ch}}}}{\left.  R(\sqrt[3]{{\gamma}
N_{\text{ch}}})\right\vert _{W_{0}}}\label{pTWdep}%
\end{equation}
where $W_{0}$ corresponds to the energy for which the interaction radius has
been computed. For p-p $W_{0}=7$ TeV if we use $R_{\text{pp}}$ from
Eq.(\ref{Rpp}), and for p-Pb $W_{0}=5.02$ TeV corresponding to $R_{\text{pPb}%
}$ of Eq.(\ref{RpPb}).

Equation (\ref{pTWdep}) implies that the  effective interaction radius at fixed
multiplicity varies with energy as $(W_0/W)^{\lambda/(2+\lambda)}\left. R \right\vert _{W_0}$. 
This is depicted in Fig.~\ref{fig:radii} where we plot $R(N_{\rm g}^{1/3})$ for three different energies.
The lowest solid (black) curve corresponds to the parametrization (\ref{Rpp}) at $W_0=7$~TeV, 
whereas two upper solid curves correspond to  (\ref{Rpp}) multiplied by the energy dependent factor
$(W_0/W)^{\lambda/(2+\lambda)}$ for $W=2.76$~TeV (red) and 0.9~TeV (blue). 

 It is however clear that the power law increase of effective interaction radius at low
energies has to tamed at some point. It is not possible in a model independent way
to find how this actually happens. Therefore we have assumed a simplistic model
that radii at all energies saturate at about  1.5~fm, a value corresponding to the CGC
prediction at 7~TeV. This is shown in Fig.~\ref{fig:radii} by dashed lines corresponding
to the sharp cut-off of the interaction radii. 
Of course the sharp cut-off is a very naive
assumption and in reality the approach to radius saturation is certainly more complicated as
suggested by the disagreement of the 0.9 TeV with fixed radius saturation hypothesis.
This issue can be also addressed by reverting the logic and by extracting the interaction radius
from the data. An attempt in this direction is briefly discussed in Ref.~\cite{Praszalowicz:2014kaa}.

\begin{figure}[h]
\centering
\includegraphics[height=6.0cm]{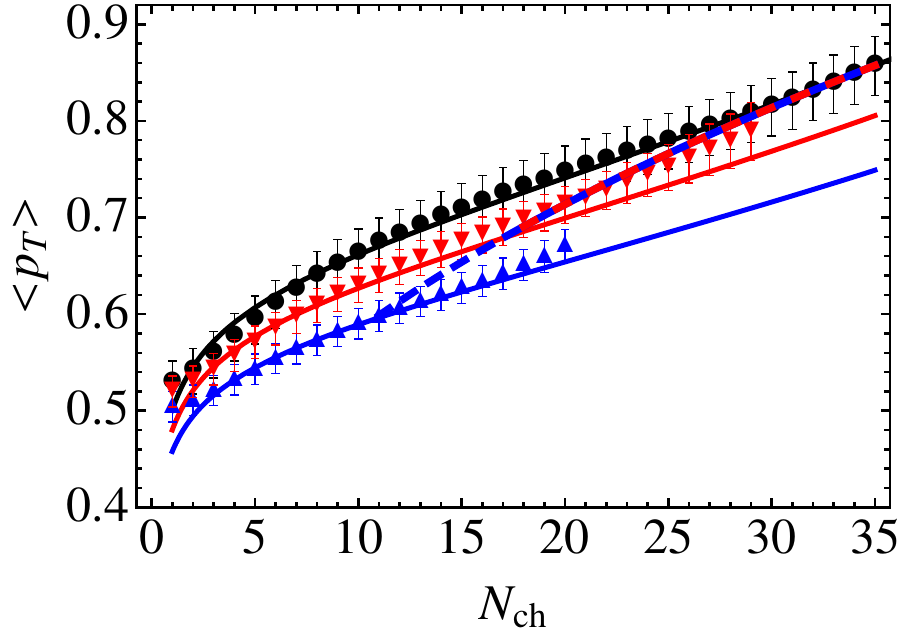}\caption{Mean $\left\langle
p_{\text{T}}\right\rangle $ in p-p collisions for three different LHC energies
7 TeV (full black circles), 2.76 TeV (full red down-triangles) and 0.9 TeV
(full blue down-triangles) together with theoretical parametrizations of
Eq.(\ref{pTWdep}). Solid lines correspond to the interaction radii $R$ shown
in Fig.~\ref{fig:radii} as the solid lines as well, whereas dashed lines show the
change in the multiplicity dependence caused by sharp $R$ saturation
shown in Fig.~\ref{fig:radii} as dashed lines.}%
\label{fig:ppwithfit}%
\end{figure}

We can now check these ideas against experiment using ALICE data on p-p, p-Pb
and Pb-Pb scattering \cite{Abelev:2013ala,Abelev:2013bla}. 
Let us first consider the case where interaction radii at different energies 
saturate at different values corresponding to the solid lines in 
Fig.~\ref{fig:radii}.

We have used the
p-p data at 7 TeV as the reference fitting to it formula (\ref{meanpTfit})
with the following result:%
\begin{equation}
\alpha=0.268\;\text{GeV},\qquad\beta=0.153\;\text{GeV}\times\text{fm,}%
\qquad\gamma=1.138.\label{pars}%
\end{equation}
One would naively expect%
\[
\gamma\simeq\frac{3}{2}\frac{1}{\Delta y}%
\]
which for ALICE pseudo-rapidity interval $\left\vert \eta\right\vert <0.3$
would give 2.5 rather than 1.138. Predictions for other two LHC energies
follow from Eq.(\ref{pTWdep}) with no other parameters. The result is plotted
in Fig.~\ref{fig:ppwithfit} and one can see good  but not perfect 
agreement with the data. One can observe that the 2.76~TeV points (red down-triangles)
at higher multiplicities tend towards the 7~TeV curve, and that two last 0.9~TeV 
points (blue up-triangles) seem to show the similar tendency. This behavior can
be attributed to the fixed value saturation of  $R$ as depicted  in Fig.~\ref{fig:radii}
by the dashed lines. The effect of fixed $R$ is shown by the dashed lines 
in Fig.~\ref{fig:ppwithfit}. One can see that 2.76~TeV data follow quite closely
the fixed saturation radius prediction starting from $N_{\rm ch} \sim  17$, whereas
the 0.9~TeV are well below the dashed line.

It is important to note at this point that the fit leading to Eq.~(\ref{pars}) is performed
over the multiplicities measured at 7~TeV that correspond to the most right (black dashed) vertical line
in Fig.~\ref{fig:radii}. One can see that the fit is driven totally by the almost linear rise
of $R$ with $N_{\rm g}^{1/3}$ with some sensitivity to the curvature before saturation,
and does not depend on the value of the saturation radius. On the contrary, lower energy data 
 in fixed saturation radius scenario,
are more sensitive both to the curvature and the value of the saturation radius for which, however,
we do not have model calculation. Therefore our analysis can be only qualitative at this
point. With more data at higher multiplicities one could make a global fit to disentangle 
the functional dependence of $R$ on multiplicity in a model independent way. 

\begin{figure}[h]
\centering
\includegraphics[height=6.0cm]{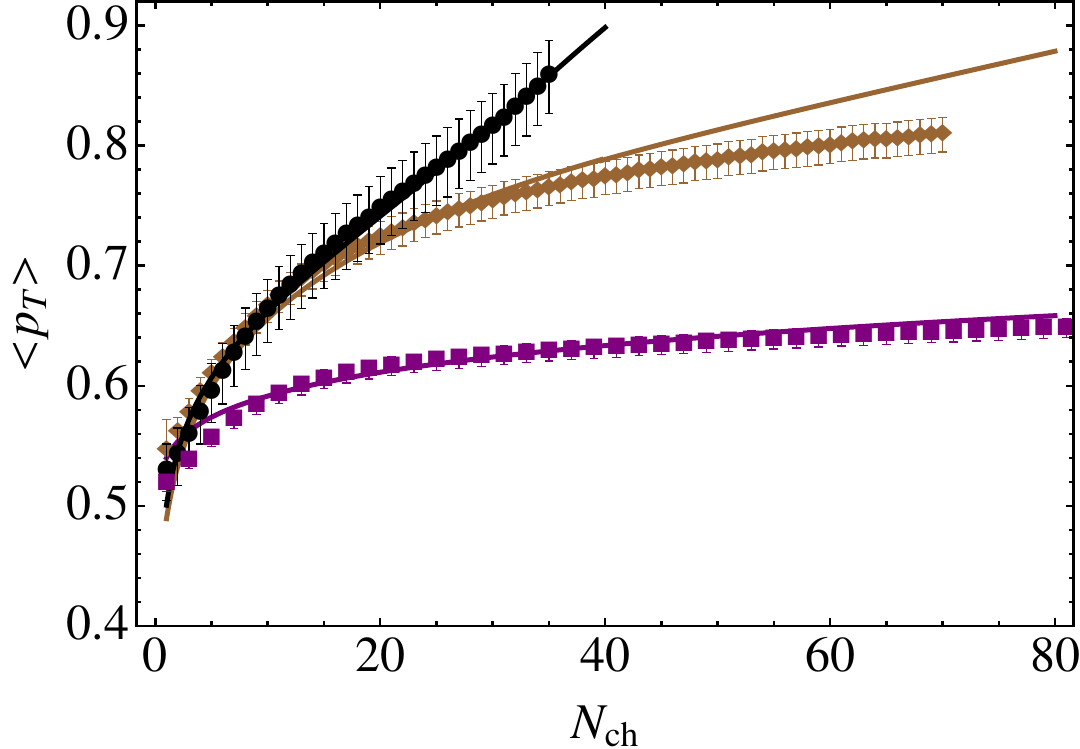}\caption{Mean
$\left\langle p_{\text{T}}\right\rangle $ in p-p collisions at 7 TeV (full
black circles), in p-Pb collisions at 5.02 TeV (full brown diamonds) and in
Pb-Pb collisions at 2.76 TeV (full purple squares) together with
parametrizations of Eqs.~(\ref{meanpTfit}) and (\ref{meanPtPbPb}).}%
\label{fig:ppPbPb}%
\end{figure}

Analogously we can calculate $\left\langle p_{\text{T}}\right\rangle $ for
p-Pb collisions using the same values of parameters (\ref{pars}) with
$R=R_{\text{pPb}}$ of Eq.(\ref{RpPb}). The result is plotted in
Fig.~\ref{fig:ppPbPb}. For comparison we also plot in Fig.~\ref{fig:ppPbPb}
$\left\langle p_{\text{T}}\right\rangle $ for p-p collisions at 7 TeV.

Finally we would like to check if mean $p_{\text{T}}$ in Pb-Pb can be also
described by formula (\ref{meanpTfit}). Unfortunately there is no calculation
of the interaction radius dependence on $dN_{\rm g}/dy$ for heavy ion collisions.
Making the plausible assumption that%

\begin{equation}
R_{\text{PbPb}}=\text{const. }\sqrt[3]{N_{\text{ch}}}\label{Pbradius}%
\end{equation}
which simply states that the saturation radius where formula (\ref{Pbradius})
should flatten is much larger than in the case of p-p and p-Pb collisions, and
should not play any role in the region where data for the latter reactions are
available. We have performed a fit to the Pb-Pb data using the following
formula%
\begin{equation}
\left\langle p_{\text{T}}\right\rangle _{\text{PbPb}}=\alpha_{\text{PbPb}%
}+\beta_{\text{PbPb}}\frac{\sqrt{N_{\text{ch}}}}{\sqrt[3]{N_{\text{ch}}}%
}\label{meanPtPbPb}%
\end{equation}
obtaining
\begin{equation}
\alpha_{\text{PbPb}}=0.43\;\text{GeV,}\qquad\beta_{\text{PbPb}}%
=0.11\;\text{GeV}.
\end{equation}
The data and the fit are also plotted in Fig.~\ref{fig:ppPbPb}.

\begin{figure}[h]
\centering
\includegraphics[height=6.0cm]{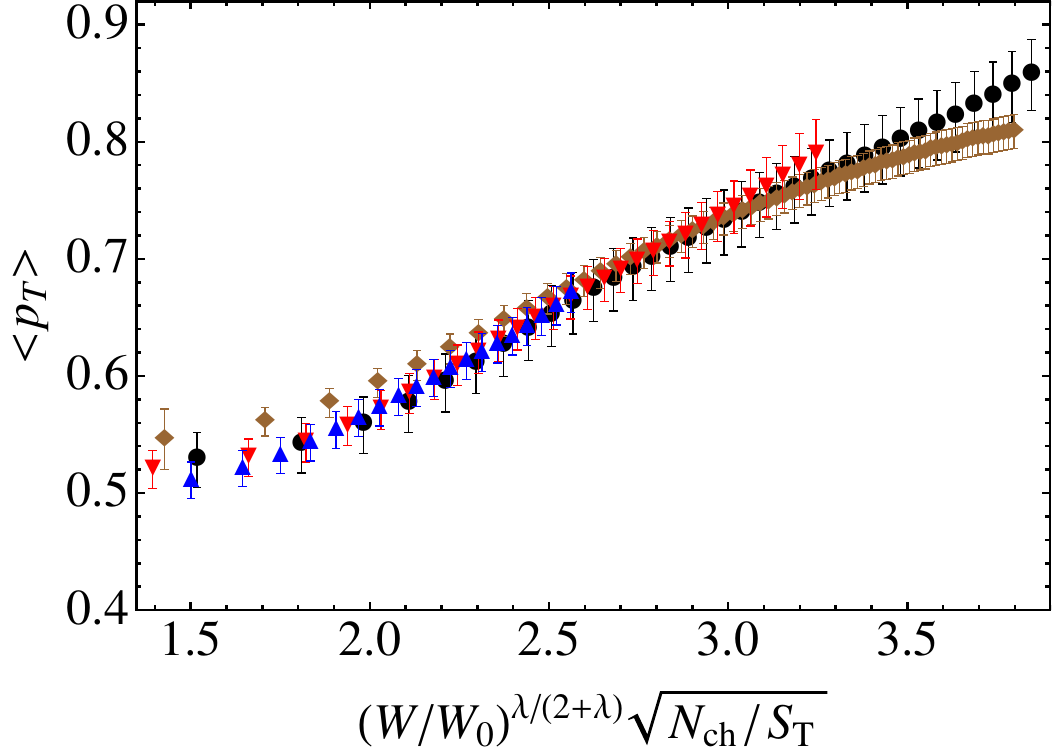}\caption{Mean
$\left\langle p_{\text{T}}\right\rangle $ in p-p collisions at 7 TeV (full
black circles), 2.76 TeV (full red down-triangles), 0.9 TeV (full blue
up-triangles) and in p-Pb collisions at 5.02 TeV (full brown diamonds) plotted
in terms of scaling variable $(W/W_{0})^{\lambda/(2+\lambda)}\sqrt
{N_{\mathrm{ch}}/ S_{\bot}}$. For p-p $W_{0}=7$~TeV and for p-Pb $W_{0}%
=5.02$~TeV.}%
\label{fig:scaled}%
\end{figure}

Yet another illustration of the mean $p_{\mathrm{T}}$ scaling is shown in
Fig.~\ref{fig:scaled} where we plot $\,\left\langle p_{\text{T}}\right\rangle
$ as a function of the scaling variable $(W/W_{0})^{\lambda/(2+\lambda)}%
\sqrt{N_{\mathrm{ch}}/S_{\bot}}$ both for p-p and p-Pb collisions. We see
quite satisfactory scaling in contrary to the claim of
Ref.~\cite{Abelev:2013bla} where the scaling variable has not been rescaled by
the energy factor $(W/W_{0})^{\lambda/(2+\lambda)}$. We cannot superimpose the
Pb-Pb data on the plot in Fig.~\ref{fig:scaled} because we do not know the
absolute normalization of $R_{\mathrm{PbPb}}$ (\ref{Pbradius}), which can be
found only by explicit calculation within the CGC effective theory.

From this simple exercise we may conclude that mean $p_{\mathrm{T}}$
dependence on charged particle multiplicity can be well described in an
approach based on the Color Glass Condensate and Geometrical Scaling. More
understanding is certainly required as far as heavy ion data are concerned,
although an an onset of GS in nuclear collisions has been already
reported~\cite{Praszalowicz:2011rm,Klein-Bosing:2014uaa}. We have established
that ALICE data on charged particle multiplicity in p-p collisions exhibit
Geometrical Scaling within a reasonable window in scaling variable $\tau$ with
exponent $\lambda=0.22$. There are some differences in the value of $\lambda$
extracted from different experiments and different reactions, however, all
results fall within a window 0.22 -- 0.32. Further studies to understand these
fine effects are clearly needed. The main finding of the present work concerns
the energy dependence of $\left\langle p_{\text{T}}\right\rangle $ which is
given by the energy dependence of the average saturation scale $\bar
{Q}_{\mathrm{s}}(W)$. Our final plot, Fig.~\ref{fig:scaled}, demonstrates very
good scaling of $\left\langle p_{\text{T}}\right\rangle $ both in p-p and p-Pb
collisions. New results at higher energies, especially in the case of p-Pb,
will provide an important test of these ideas.

We have also argued that the interaction radii at different energies should
for large multiplicities converge to some fixed value. Such tendency is
clearly seen at 2.76~TeV and presumably at 0.9~TeV at multiplicities
 above 20. Our simplistic sharp cut-off model fails to describe 0.9~TeV
 data, but that could be presumably cured not affecting the other energies,
 by allowing for a somewhat
 larger value of the saturation radius and careful modeling of the curvature
 before saturation.
We find it quite unexpected that such simple observable as 
$ \left\langle p_{\text{T}}\right\rangle$ can provide such nontrivial information on
the energy and multiplicity behavior of the interaction radius.

\section*{Acknowledgements}

This research of M.P. has been supported by the Polish NCN grant 2011/01/B/ST2/00492.  The research of L.M. is supported under DOE Contract No. DE-AC02-98CH10886.


\section*{References}

\begin{thebibliography}{99}                                                                                               %



\bibitem {Abelev:2013ala}B.~B.~Abelev \textit{et al.} [ALICE Collaboration],
Eur.\ Phys.\ J.\ C \textbf{73} (2013) 2662 [arXiv:1307.1093 [nucl-ex]].




\bibitem {Abelev:2013bla}B.~B.~Abelev \textit{et al.} [ALICE Collaboration],
Phys.\ Lett.\ B \textbf{727} (2013) 371 [arXiv:1307.1094 [nucl-ex]].




\bibitem {McLerran:2013oju}L.~McLerran, M.~Praszalowicz and B.~Schenke,
Nucl.\ Phys.\ A \textbf{916} (2013) 210 [arXiv:1306.2350 [hep-ph]].




\bibitem {McLerran:2010ex}L.~McLerran and M.~Praszalowicz,
Acta Phys.\ Pol.\ B \textbf{41} (2010) 1917 [arXiv:1006.4293 [hep-ph]], and
Acta Phys.\ Pol.\ B \textbf{42} (2011) 99 [arXiv:1011.3403 [hep-ph]].


\bibitem {CMS}V.~Khachatryan \textit{et al.} [CMS Collaboration],
JHEP \textbf{1002} (2010) 041 [arXiv:1002.0621 [hep-ex]];\newline
V.~Khachatryan \textit{et al.} [CMS Collaboration],
Phys.\ Rev.\ Lett.\ \textbf{105} (2010) 022002 [arXiv:1005.3299 [hep-ex]].



\bibitem {Praszalowicz:2012zh}M.~Praszalowicz and T.~Stebel,
JHEP \textbf{1303} (2013) 090 [arXiv:1211.5305 [hep-ph]].


\bibitem {HERAdata}C.~Adloff \textit{et al.} [H1 Collaboration],
{Eur.\ Phys.\ J.\ C} \textbf{21} (2001) 33 [arXiv:hep-ex/0012053];\newline
S.~Chekanov \textit{et al.} [ZEUS Collaboration],
{Eur.\ Phys.\ J.\ C} \textbf{21} (2001) 443 [arXiv:hep-ex/0105090].




\bibitem {Bzdak:2013zma}A.~Bzdak, B.~Schenke, P.~Tribedy and R.~Venugopalan,
Phys.\ Rev.\ C \textbf{87} (2013) 064906 [arXiv:1304.3403 [nucl-th]].


\bibitem {MLV}L.~D.~McLerran and R.~Venugopalan,
Phys.\ Rev.\ \textbf{D49} (1994) 2233 [hep-ph/9309289], \ Phys. Rev.
\textbf{D49} (1994) 3352 [hep-ph/9311205], \ and Phys.\ Rev.\ \textbf{D50}
(1994) 2225 [hep-ph/9402335].



\bibitem {Fukushima:2011ca}K.~Fukushima,
Acta Phys.\ Pol.\ B \textbf{42} (2011) 2697 [arXiv:1111.1025 [hep-ph]].


\bibitem {Kharzeev:2000ph}D.~Kharzeev and M.~Nardi,
Phys.\ Lett.\ B \textbf{507}, 121 (2001) [arXiv:nucl-th/0012025].




\bibitem {Kharzeev:2001gp}D.~Kharzeev and E.~Levin,
Phys.\ Lett.\ B \textbf{523} (2001) 79 [arXiv:nucl-th/0108006].




\bibitem {Kharzeev:2002ei}D.~Kharzeev, E.~Levin and M.~Nardi,
Nucl.\ Phys.\ A \textbf{730} (2004) 448 [Erratum-ibid.\ A \textbf{743}, 329
(2004)] [arXiv:hep-ph/0212316],
Nucl.\ Phys.\ A \textbf{747} (2005) 609 [arXiv:hep-ph/0408050],
Phys.\ Rev.\ C \textbf{71} (2005) 054903 [arXiv:hep-ph/0111315].

\bibitem{Kovchegov:1998bi}
  Y.~V.~Kovchegov and A.~H.~Mueller,
  Nucl.\ Phys.\ B {\bf 529} (1998) 451
  [hep-ph/9802440].

\bibitem{Dumitru:2001ux}
  A.~Dumitru and L.~D.~McLerran,
  Nucl.\ Phys.\ A {\bf 700} (2002) 492
  [hep-ph/0105268].





\bibitem {Stasto:2000er}A.~M.~Stasto, K.~J.~Golec-Biernat and J.~Kwiecinski,
Phys.\ Rev.\ Lett.\ \textbf{86} (2001) 596 [hep-ph/0007192].




\bibitem {Praszalowicz:2011tc}M.~Praszalowicz,
Phys.\ Rev.\ Lett.\ \textbf{106} (2011) 142002 [arXiv:1101.0585 [hep-ph]].


\bibitem {Praszalowicz:2011rm}M.~Praszalowicz,
Acta Phys.\ Pol.\ B \textbf{42} (2011) 1557 [arXiv:1104.1777 [hep-ph]], and
in \emph{Proceedings of the 47th Rencontres de Moriond, La Thuile, 2012}, p.
265 [arXiv:1205.4538 [hep-ph]].


\bibitem {jimwlk}J. Jalilian-Marian, A. Kovner, A. Leonidov and H. Weigert,
Nucl.\ Phys.\ B \textbf{504} (1997) 415 [hep-ph/9701284] and Phys. Rev.
\textbf{D59} (1998) 014014 [hep-ph/9706377]; \newline E. Iancu, A. Leonidov,
and L. D. McLerran, Nucl. Phys. \textbf{A692} (2001) 583 [hep-ph/0011241];
\newline E. Ferreiro, E. Iancu, A. Leonidov and L. D. McLerran, Nucl. Phys.
\textbf{A703} (2002) 489 [hep-ph/0109115].

\bibitem {BK}I.~Balitsky,
Nucl.\ Phys.\ B \textbf{463} (1996) 99 [hep-ph/9509348]; \newline
Y.~V.~Kovchegov,
Phys.\ Rev.\ \textbf{D60} (1999) 034008 [hep-ph/9901281], and
Phys.\ Rev.\ D \textbf{61} (2000) 074018 [hep-ph/9905214].



\bibitem {Munier:2003vc}S.~Munier and R.~B.~Peschanski,
Phys.\ Rev.\ Lett.\ \textbf{91} (2003) 232001 [hep-ph/0309177], and
Phys.\ Rev.\ D \textbf{69} (2004) 034008 [hep-ph/0310357].




\bibitem {Gelis:2009wh}F.~Gelis, T.~Lappi and L.~McLerran,
Nucl.\ Phys.\ A \textbf{828} (2009) 149 [arXiv:0905.3234 [hep-ph]].




\bibitem {Dumitru:2010iy}A.~Dumitru, K.~Dusling, F.~Gelis, J.~Jalilian-Marian,
T.~Lappi and R.~Venugopalan,
Phys.\ Lett.\ B \textbf{697} (2011) 21 [arXiv:1009.5295 [hep-ph]].


\bibitem {Francuz}A. Francuz and M. Praszalowicz in preparation.



\bibitem {Klein-Bosing:2014uaa}C.~Klein-B{\"o}sing and L.~McLerran,
Phys.\ Lett.\ B \textbf{734} (2014) 282 [arXiv:1403.1174 [nucl-th]].

\bibitem{Praszalowicz:2013fsa}
  M.~Praszalowicz,
  Phys.\ Lett.\ B {\bf 727} (2013) 461
  [arXiv:1308.5911 [hep-ph]].




\bibitem {Aamodt:2009aa}K. Aamodt \textit{et al.} [ALICE Collaboration],
Eur.\ Phys.\ J.\ C \textbf{65} (2010) 111 [arXiv:0911.5430 [hep-ex]].

\bibitem{Khachatryan:2010us}
  V.~Khachatryan {\it et al.}  [CMS Collaboration],
  Phys.\ Rev.\ Lett.\  {\bf 105} (2010) 022002
  [arXiv:1005.3299 [hep-ex]].
 
\bibitem{Tribedy:2010ab}
  P.~Tribedy and R.~Venugopalan,
  Nucl.\ Phys.\ A {\bf 850} (2011) 136
   [Erratum-ibid.\ A {\bf 859} (2011) 185]
  [arXiv:1011.1895 [hep-ph]].

  
\bibitem{Praszalowicz:2014kaa}
  M.~Praszalowicz,
  arXiv:1410.5220 [hep-ph].

\end{thebibliography}
\end{document}